\documentclass[aps,pra,twocolumn,floatfix]{revtex4}

\usepackage{psfrag,graphicx}
\usepackage{hyperref}
\usepackage{amsmath,amssymb}
\newcommand{\ket}[1]{|#1\rangle}

\def\>{\rangle}
\def\<{\langle}

\begin{document}

\numberwithin{equation}{section}

\title{Effects of self-phase modulation on weak nonlinear optical quantum gates}

\author{Pieter Kok}\email{p.kok@sheffield.ac.uk}
\affiliation{Department of Physics and Astronomy, University of
  Sheffield, Hounsfield Road, Sheffield, S3 7RH, UK}  
\affiliation{Department of Materials, University of Oxford, Parks Road,
  Oxford, OX1 3PU, UK}

\begin{abstract}
\noindent A possible two-qubit gate for optical quantum computing is
the parity gate based on the weak Kerr effect. Two photonic qubits
modulate the phase of a coherent state, and a quadrature measurement
of the coherent state reveals the parity of the two qubits without
destroying the photons. This can be used to create so-called cluster
states, a universal resource for quantum computing. Here, the effect
of self-phase modulation on the parity gate is studied, introducing
generating functions for the Wigner function of a modulated coherent
state. For materials with non-{\sc eit}-based Kerr nonlinearities,
there is typically a self-phase modulation that is half the magnitude
of the cross-phase modulation. Therefore, this effect cannot be
ignored. It is shown that for a large class of physical
implementations of the phase modulation, the quadrature measurement
cannot distinguish between odd and even parity. Consequently, weak
nonlinear parity gates must be implemented with physical systems where
the self-phase modulation is negligable. 
\end{abstract}


\maketitle

\section{Introduction}

\noindent
Linear optical quantum computing with photonic qubits has generated
considerable interest in recent years \cite{knill01,kok07}. However,
it has become clear that from a scaling perspective, some optical
nonlinearity is extremely desirable. This can be achieved in a number
of ways, either by using the coupling of photons with matter qubits
\cite{barrett05,lim05,lim06}, quantum Zeno gates \cite{franson04} or
by using optical nonlinearities
\cite{barrett05b,nemoto04,spiller06}. To date, analysis of weak
nonlinear optical gates has not taken into account self-phase
modulation, mainly because the envisaged implementation is materials
with an electromagnetically induced transparency ({\sc eit}). It is
known that in such systems a large cross-phase modulation (Kerr
nonlinearity) can be achieved without any self-phase modulation
\cite{schmidt96}. However, in general nonlinear optical media
self-phase modulation effects are present, and are typically of the
order of half the cross-phase modulation. This is known as weak-wave
retardation \cite{chiao66,boyd03}.   

In this paper, we analyse the effects of self-phase modulation on the
operation of the weak nonlinear optical parity gate, which was first
introduced by Barrett et al. \cite{barrett05b}. To this end, the
classical and quantum theory of materials with third-order optical
nonlinearities is reviewed, and we develop a description of the
self-phase modulation of a coherent state in terms of its Wigner
function and the corresponding marginal probability distributions for
the quadratures. We then give a modified, more realistic description
of the parity gate, and show that a typical amount of self-phase
modulation destroys the distinguishability between even and odd
parity. We conclude that the nonlinearities in weak nonlinear optical
parity gates must have negligable self-phase modulation.  

\begin{figure}[t]
\begin{center}
\begin{psfrags}
 \psfrag{a}{\large $\theta$}
 \psfrag{b}{$-$\large $\theta$}
 \psfrag{c}{$2\vartheta(x)$}
 \psfrag{d}{\large $\ket{\alpha}$}
 \psfrag{e}{\large $\ket{\psi_1}$}
 \psfrag{f}{\large $\ket{\psi_2}$}
  \includegraphics[width=77mm]{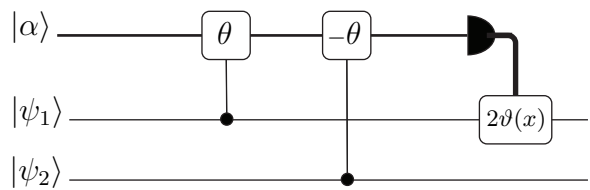}
 \end{psfrags}
 \end{center}
 \caption{Controlled phase shifts $\theta$ and $-\theta$ on a coherent state $|\alpha\rangle$, followed by an $x$-quadrature measurement induce a deterministic and nondestructive parity measurement. A corrective phase shift of $2\vartheta(x)$ must be applied when the measurement outcome $x$ indicates odd parity. The conditional phase shifts are generated by a cross-Kerr nonlinearity.}
 \label{EasyAs123}
\end{figure}

The original optical parity gate based on conditional phase shifts
without self-phase modulation effects can be summarized as follows
\cite{barrett05b}: Two single-photon qubits each couple to an optical
mode in a coherent state $|\alpha\rangle$, such that a phase shift is
induced in the coherent state whenever the qubit is in the logical
state $|1\rangle$. Moreover, the first qubit induces a controlled
phase shift $\theta$, while the second qubit induces a controlled
phase shift $-\theta$ (see Fig.~\ref{EasyAs123}). The input state
$(c_{00}|00\rangle + c_{01}|01\rangle + c_{10}|10\rangle +
c_{11}|11\rangle)|\alpha\rangle$ transforms into the output state 
\begin{equation}\nonumber
	c_{00}|00\rangle|\alpha\rangle + c_{01}|01\rangle|\alpha e^{-i\theta}\rangle + c_{10}|10\rangle|\alpha e^{i\theta}\rangle + c_{11}|11\rangle|\alpha\rangle .
\end{equation}
If we assume that $\alpha\in\mathbb{R}$, then an $\hat{x}$ quadrature
measurement of the coherent state will project the qubit states either
onto the even parity subspace ($c_{00}|00\rangle + c_{11}|11\rangle$),
or onto the odd parity subspace $(c_{01} |01\rangle +
e^{2i\vartheta(x)} c_{10} |10\rangle)$. When the measurement outcome
$x$ indicates projection onto the odd subspace, a known corrective
phase shift $2\vartheta(x)$ must be applied. Hence this is a
deterministic parity gate that can be used to build cluster states
\cite{barrett05,browne05}, which are a universal resource for quantum
computing. The requirements for this gate to work with sufficiently
high fidelity is that $|\langle\alpha|\alpha e^{i\theta}\rangle| \ll
1$, or $\alpha\theta^2 \gtrsim 1$. In the modified protocol of Spiller
et al., \cite{spiller06} the requirement becomes
$\alpha\theta>1$. This scaling is important in practical
implementations, where we require $\alpha$ to be reasonably small
\cite{barrett06}. The coherent state is often called the {\em bus},
and the conditional phase shifts are implemented with optically active
materials such as Kerr nonlinearities.  

\section{Theory of Kerr nonlinearities}

\noindent
The weak nonlinear optical parity gate for photonic qubits uses the
optical Kerr effect. In this section we review the classical and
quantum theory of the third-order nonlinear interaction that leads to
the Kerr effect, and we present a convenient way to describe the Kerr
effect on coherent states using Wigner functions. 

\subsection{Classical and quantum theory}

\noindent
A detailed review of the classical theory of the Kerr effect can be
found in Boyd \cite{boyd03}, and here we give a brief description of
the relevant physics. For a lossless and dispersionless nonlinear
medium, the optical response can be expressed as a power series in the
electric field of the polarization field in the medium: 
\begin{equation}
	P(t) = \chi^{(1)} E(t) + \chi^{(2)} E(t)^2 + \chi^{(3)} E(t)^3\; . 
	\label{eq:polarization}
\end{equation}
where $P(t) = \sum_k P(\omega_k) e^{-i\omega_k t} + {\rm c.c.}$ and $E(t) = \sum_n E(\omega_n) e^{-i\omega_n t} + {\rm c.c.}$ are multi-mode expansions of the polarization and electric field. Here, we are particularly interested in the third order term $P^{(3)}(t) = \chi^{(3)} E(t)^3$. The polarizability and the electric fields are of course vector fields with three spatial components, and in general we should write 
\begin{equation}
	P_i^{(3)}(t) = \sum_{jkl} \chi_{ijkl}^{(3)} E_j(t) E_k(t) E_l(t)\; .
	\label{eq:chi-tensor}
\end{equation}
The nonlinear susceptibility $\chi_{ijkl}^{(3)}$ is then a fourth-rank
tensor with 81 independent components. However, most materials are
highly symmetric, and the actual number of independent components is
far less. Here, we take $\chi^{(3)}$ to be independent of the
orientation of the fields, and treat the electric field and the
polarization as scalar quantities. 

The nonlinear medium required for the optical parity gate couples two
modes, $a$ and $b$, with possibly different frequencies $\omega_a$ and
$\omega_b$. The frequencies may be identical, as long as the modes are
distinguishable (e.g., they have different spatial directions). When
we substitute this two-mode expansion of the electric field into
Eq.~(\ref{eq:polarization}) and collect all terms that are
proportional to $e^{-i\omega_a t}$, the third order of the
polarization field consists of two terms, namely the {\em cross-phase}
modulation ({\sc xpm}) 
\begin{equation}
	P^{(3)}_{XPM}(\omega_a) = 6 \chi^{(3)} E(\omega_a) |E(\omega_b)|^2\; ,
	\label{eq:xpm}
\end{equation}
and the {\em self-phase} modulation ({\sc spm})
\begin{equation}
	P^{(3)}_{SPM}(\omega_a) = 3 \chi^{(3)} E(\omega_a) |E(\omega_a)|^2\; .
	\label{eq:spm}
\end{equation}
Here, we have suppressed the time dependence. In {\sc xpm}, the
polarization of mode $a$ depends on the intensity of the field in mode
$b$, while in {\sc spm} the polarization in mode $a$ depends on the
intensity in mode $a$. Notice the relative factor of two in the
magnitude of the phase modulations. This means that for general
isotropic nonlinear media the self-phase modulation is typically half
the size of the cross-phase modulation.  

The index of refraction $n$ for a specific mode in a medium can be
defined in terms of an effective susceptibility $\chi_{\rm eff}$ such
that $n^2 = 1+4\pi \chi_{\rm eff}$. Taking into account the
third-order polarizability $P$ (and assuming a vanishing second-order
susceptibility) we have for mode $a$: 
\begin{equation}
	\chi_{\rm eff} \equiv \chi^{(1)} + 3\chi^{(3)} |E(\omega_a)|^2 + 6\chi^{(3)} |E(\omega_a)|^2\; ,
	\label{eq:chi-eff}
\end{equation}
In other words, there will be a phase shift in mode $a$ that is
proportional not only to the regular phase shift ($\chi^{(1)}$), but
also proportional to the intensity $|E|^2$ in mode $a$ and in mode
$b$.  

Quantum mechanically, phase shifts are generated by the number
operator of a given optical mode: 
\begin{equation}
	e^{i\varphi\hat{a}^{\dagger}\hat{a}}\, \hat{a}\, e^{-i\varphi\hat{a}^{\dagger}\hat{a}} = \hat{a}\, e^{-i\varphi}\; .
	\label{eq:quantum-phase}
\end{equation}
When the phase shift of mode $a$ is proportional to the intensity of
the field in mode $b$, one would expect the phase shift generator
$\hat{n}_a = \hat{a}^{\dagger}\hat{a}$ to be multiplied by the number
operator of mode $b$. Indeed, the {\sc xpm} and {\sc spm} effects are
described by the interaction Hamiltonians 
\begin{equation}
	H_{XPM} = \theta\, \hat{a}^{\dagger}\hat{a}\, \hat{b}^{\dagger}\hat{b} = \theta\, \hat{n}_a \hat{n}_b\; ,
	\label{eq:xpm-hamiltonian}
\end{equation}
and 
\begin{equation}
	H_{SPM} = \phi_a (\hat{a}^{\dagger}\hat{a})^2 + \phi_b (\hat{b}^{\dagger}\hat{b})^2 = \phi_a\, \hat{n}_a^2 + \phi_b\, \hat{n}_b^2\; .
 	\label{eq:spm-hamiltonian}
\end{equation}
The coupling constants $\theta$ and $\phi_j$ ($j=a,b$) are
proportional to $6\chi^{(3)}$ and $3\chi^{(3)}$, respectively. In the
remainder of this paper we set $\phi_a = \phi_b = \phi$, and towards
the end we shall set $\theta = 2\phi$. We note a couple of things:
First, the signs of $\theta$ and $\phi$ must be identical, since they
are both proportional to $\chi^{(3)}$. This will become important when
we analyze the weak nonlinear parity gate. And secondly, since
$[H_{XPM},H_{SPM}]=0$, we can treat the self-phase modulation
independent from the cross-phase modulation.  

Given the interaction Hamiltonian $H_{SPM}$, we can calculate the effect of {\sc spm} on a coherent state:
\begin{equation}
	e^{-i\phi \hat{n}^2} |\alpha\rangle = e^{-|\alpha|^2/2}\, \sum_{n=0}^{\infty} \frac{\alpha^n\, e^{-i\phi n^2}}{\sqrt{n!}} |n\rangle\; .
	\label{eq:spm-coh}
\end{equation}
However, the oft-needed inner products $\langle x_{\lambda}|e^{-i\phi
  \hat{n}^2} |\alpha\rangle$ and $\langle \beta|e^{-i\phi \hat{n}^2}
|\alpha\rangle$ cannot be evaluated analytically. To find the effect
of {\sc spm} on measured quantities, we should instead transform the
corresponding observables using $\exp(-i H_{SPM})$.

The operator transformations for the creation and annihilation
operators of the field modes, given the {\sc xpm} and {\sc spm}
interaction Hamiltonians (denoted by $\varphi A$), can formally be
written as 
\begin{equation}
	e^{i\varphi A}\, \hat{a} e^{-i\varphi A} = \hat{a} + i\varphi [A,\hat{a}] + \frac{(i\varphi)^2}{2!} [A,[A,\hat{a}]] + \ldots
	\label{eq:formal-operator-transform}
\end{equation}
It is straightforward to show that 
\begin{eqnarray}
	e^{iH_{XPM}}\, \hat{a} e^{-iH_{XPM}} &=& \hat{a} e^{-i\theta\hat{n}_b} \cr
	e^{iH_{SPM}}\, \hat{a} e^{-iH_{SPM}} &=& e^{-i\theta(2\hat{n}_a+1)} \hat{a} \cr
	e^{iH_{SPM}}\, \hat{a}^{\dagger} e^{-iH_{SPM}} &=& \hat{a}^{\dagger} e^{i\theta(2\hat{n}_a+1)}\; .
	\label{eq:bogs}
\end{eqnarray}
For the {\sc spm} case, the operators are ordered such that all
creation operators are on the left, the phase operators
$\exp[-i\phi\hat{n}_a]$ are in the center, and the annihilation
operators are on the right. This will be convenient later, when we
evaluate expectation values with respect to coherent states. 

The quadrature operators of the electromagnetic field can be defined
as \cite{barnett} 
\begin{equation}
	\hat{x}_{\lambda} = \frac{1}{\sqrt{2}}\left( \hat{a}\, e^{-i\lambda} + \hat{a}^{\dagger}\, e^{i\lambda} \right) .
	\label{eq:quadrature}
\end{equation}
We can construct the canonical momentum to this operator as
$\hat{x}_{\lambda+\pi/2}$, since
$[\hat{x}_{\lambda},\hat{x}_{\lambda+\pi/2}]=i$. The effect of {\sc
  spm} on the quadrature operator results in a transformed operator
$\hat{x}_{\lambda}'$ 
\begin{equation}
	\hat{x}_{\lambda}' = \frac{1}{\sqrt{2}}\left( e^{-i\lambda-i\phi(2\hat{n}+1)}\, \hat{a} + \hat{a}^{\dagger}\, e^{i\lambda+i\phi(2\hat{n}+1)} \right) .
	\label{eq:trans-quad}
\end{equation}
The expression in Eq.~(\ref{eq:trans-quad}) allows us to evaluate the
moments of $\hat{x}_{\lambda}'$, and consequently we can calculate the
mean, variance, skewness, and kurtosis of the probability distribution
of measurement outcomes for $\hat{x}_{\lambda}'$.

\subsection{Wigner function of self-phase modulated coherent state}

\noindent
In the weak nonlinear parity gate, the {\sc spm} affects the coherent
state that acts as the bus mode. The qubits are single photons, and
are affected only in a trivial way by the {\sc spm}, in that it
induces a known phase shift in the qubit states. In this section, we
derive the Wigner function for a self-phase modulated coherent state. 

Any single-mode state of the electromagnetic field can be written in a
photon number expansion $|\Psi\rangle = \sum_n A_n
|n\rangle$. Furthermore, we can insert a resolution of the identity in
terms of coherent states such that  
\begin{equation}
	|\Psi\rangle = \sum_{n=0}^{\infty} A_n |n\rangle = \sum_{n=0}^{\infty} \int \frac{d^2\beta}{\pi} A_n |\beta\rangle\langle\beta|n\rangle\; .
	\label{eq:}
\end{equation}
For $|\Psi\rangle$ a coherent state experiencing {\sc spm}, we found that 
\begin{equation}
	A_n = e^{-|\alpha|^2/2} \frac{\alpha^n}{\sqrt{n!}}\, e^{-i\phi n^2}\; .
	\label{eq:an}
\end{equation}
Using the inner product $\langle\beta|n\rangle = e^{-|\beta|^2/2} \beta^{*n}/\sqrt{n!}$ we find
\begin{equation}
	|\Psi\rangle = \sum_{n=0}^{\infty} \int \frac{d^2\beta}{\pi} e^{-(|\alpha|^2+|\beta|^2)/2} \frac{(\alpha\beta^*)^n}{n!} e^{-i\phi n^2} |\beta\rangle\; .
\end{equation}
The factor $e^{-i\phi n^2}$ can be written as a power series $\sum_k
(-i\phi)^k n^{2k}/k!$: 
\begin{equation}
	|\Psi\rangle = \sum_{k,n=0}^{\infty} \int \frac{d^2\beta}{\pi} e^{-(|\alpha|^2+|\beta|^2)/2} \frac{(\alpha\beta^*)^n}{n!} \frac{(-i\phi)^k n^{2k}}{k!} |\beta\rangle\; .
\end{equation}
This expansion allows us to remove the factor $n^{2k}$ by applying the
differential operator $(\alpha\partial_{\alpha})^{2k}$:
\begin{equation}
	|\Psi\rangle = e^{-|\alpha|^2/2} \sum_{k=0}^{\infty} \frac{(-i\phi)^k (\alpha\partial_{\alpha})^{2k}}{k!}  \int \frac{d^2\beta}{\pi} e^{-|\beta|^2/2+\alpha\beta^*} |\beta\rangle\; .
\end{equation}
Rewriting the power series in $k$ as an exponential, we have
\begin{equation}
	|\Psi\rangle = e^{-|\alpha|^2/2}\, U_{\alpha}(\phi) \int \frac{d^2\beta}{\pi} e^{-|\beta|^2/2+\alpha\beta^*} |\beta\rangle\; ,
\end{equation}
where 
\begin{equation}
	U_{\alpha}(\phi) = \exp\left[ -i\phi \left( \alpha\partial_{\alpha} \right)^2 \right] .
\end{equation}

The Wigner function of this state is constructed according to ($\hbar=1$)
\begin{equation}
	W(q,p) = \frac{1}{\pi} \int_{-\infty}^{\infty} dx\, e^{-2ipx} \langle q-x |\Psi\rangle \langle\Psi|q+x\rangle\; ,
	\label{eq:wigner}
\end{equation}
and the marginal probability distribution of the $q$ quadrature is given by
\begin{equation}
	P(q) = \int_{-\infty}^{\infty} W(q,p) dp \; .
	\label{eq:marginal}
\end{equation}
To calculate the Wigner function, we first evaluate $\langle x|\Psi\rangle$,
using  $\langle x|\alpha\rangle = \pi^{-1/4}
\exp[-\frac{1}{2}(x-\sqrt{2}\alpha)^2 +
\frac{1}{2}\alpha(\alpha-\alpha^*)]$ \cite{barnett}: 
\begin{eqnarray}
	\langle x|\Psi\rangle &=& e^{-\frac{1}{2}|\alpha|^2} U_{\alpha}(\phi) \int \frac{d^2\beta}{\pi\sqrt[4]{\pi}} e^{-\frac{1}{2}x^2+\sqrt{2}\beta x+\alpha\beta^*-|\beta|^2-\frac{1}{2}\beta^2} \cr
	&=& e^{-\frac{1}{2}|\alpha|^2} U_{\alpha}(\phi)\, \frac{e^{-\frac{1}{2}x^2+\sqrt{2}\alpha x - \frac{1}{2}\alpha^2}}{\sqrt[4]{\pi}} \cr
	&\equiv& e^{-\frac{1}{2}|\alpha|^2} U_{\alpha}(\phi)\, G_{\alpha}(x)\; ,
	\label{eq:generating}
\end{eqnarray}
where we defined $G_{\alpha}(x)$ as a generating function for the
Wigner function such that 
\begin{eqnarray}
	W(q,p) &=& \frac{e^{-|\alpha|^2}}{\pi}\, U_{\beta}(\phi) U_{\gamma}^{\dagger}(\phi) \\ && \times \left. \int dx\, e^{-2ipx} G_{\beta}(q-x) G_{\gamma}^*(q+x) \right|_{\beta,\gamma = \alpha} . \nonumber
	\label{eq:wigner-generating}
\end{eqnarray}
The integral over $x$ can be evaluated to yield
\begin{equation}
	W(q,p) = \left. \frac{e^{-|\alpha|^2}}{\pi}\, U_{\beta}(\phi) U_{\gamma}^{\dagger}(\phi)\, K_{\beta,\gamma}(q,p) \right|_{\beta,\gamma=\alpha} ,
\end{equation}
with the generating function
\begin{equation}	
	K_{\beta,\gamma}(q,p) = e^{-p^2-q^2+i\sqrt{2}(\beta-\gamma^*)p+\sqrt{2}(\beta+\gamma^*)q-\beta\gamma^*} . 
	\label{eq:wigner-generating2}
\end{equation}
The marginal probability distribution over the $q$ quadrature for the
{\sc spm} coherent state then becomes 
\begin{equation}
	P(q) = \left.  \frac{e^{-|\alpha|^2}}{\sqrt{\pi}}\, U_{\beta}(\phi) U_{\gamma}^{\dagger}(\phi)\, L_{\beta,\gamma}(q)\right|_{\beta,\gamma=\alpha} ,
	\label{eq:marprob}
\end{equation}
with the generating function 
\begin{equation}
	L_{\beta,\gamma}(q) = e^{-q^2+\sqrt{2}(\beta+\gamma^*)q - \frac{1}{2}(\beta-\gamma^*)^2 - \beta\gamma^*} .
\end{equation}
The Wigner function and the marginal probability distribution can be
found by applying the differential operators $U_{\alpha}$ to the
respective generating functions. In the case of the weak nonlinear
parity gate, we are interested in the case where $\alpha\theta^2
\simeq 1$, and a first-order expansion of $U_{\alpha}$ will not be
sufficient to properly evaluate the marginal probability
distribution. Instead, it has to be evaluated numerically.

\section{The parity gate with self-phase modulation}

\noindent
The original weak nonlinear parity gate as described above is
invariant under {\sc spm}. This is because the coupling constant for
{\sc spm} and {\sc xpm} have the same sign, and in the two successive
stages of the protocol two conditional phase shifts $\theta$ with
opposite signs are used. Unfortunately, it is generally not possible
to change the sign of the conditional phase shift. The nonlinear
susceptibility $\chi^{(3)}$ is a material constant that cannot readily
be changed. Using different materials with opposite nonlinearities
seems highly impractical. In the case of {\sc eit}, the phase shift is
proportional to the detuning of the qubit mode with the relevant
transition in the {\sc eit} medium. Since both the frequency of the
single photon and the transition frequency are fixed, it is not
possible to switch the sign of the nonlinearity in the weak nonlinear
parity gate. 

However, it is possible to construct the parity gate with two
identical conditional phase shifts. Let the input state of the two
qubits and the coherent bus again be given by $(c_{00}|00\rangle +
c_{01}|01\rangle + c_{10}|10\rangle +
c_{11}|11\rangle)|\alpha\rangle$. The two coherent phase shifts now
generate the state 
\begin{equation}\nonumber
	c_{00}|00\rangle|\alpha\rangle + (c_{01}|01\rangle + c_{10}|10\rangle)|\alpha e^{i\theta}\rangle + c_{11}|11\rangle|\alpha e^{2i\theta}\rangle .
	\label{eq:rotated-parity}
\end{equation}
If $\alpha\in\mathbb{R}$, instead of measuring the $x$ quadrature, we measure $\hat{x}_{\lambda}$, where we now have to choose $\lambda$ such that 
\begin{equation}
	\langle\alpha|\hat{x}_{\lambda}|\alpha\rangle = \langle\alpha e^{2i\theta}|\hat{x}_{\lambda}|\alpha e^{2i\theta}\rangle .
	\label{eq:lambda}
\end{equation}
This leads to the requirement $\cos\lambda = \cos(2\theta-\lambda)$,
which is satisfied for $\lambda = \theta$. Constructed this way, the
parity gate is rotated by an angle $\theta$ in phase space, and the
roles of even and odd parity are reversed, in that the corrective
phase shift is now applied to the even parity outcome. The variance is
not affected by the rotation, and the distinguishability requirement
is the same as in the original setup. 

When the weak nonlinear parity gate is constructed with two identical
conditional phase shifts, the {\sc spm} no longer cancels, 
and instead of evaluating the expectation values of
$\hat{x}_{\lambda}$, we need to work with $\hat{x}_{\lambda}'$ defined
in Eq.~(\ref{eq:trans-quad}). The expectation values of
$\hat{x}_{\lambda}'$ with respect to $|\alpha\rangle$ and $|\alpha
e^{2i\theta}\rangle$ will be different from the expectation values of
$\hat{x}_{\lambda}$, and we have to recalculate the value of $\lambda$
for which the even parity contributions overlap. For a coherent state
with $\alpha = r\, e^{i\xi}$ the mean is 
\begin{equation}
	\langle\alpha|\hat{x}_{\lambda}'|\alpha\rangle = \sqrt{2}r\,
	e^{-r^2(1-\cos 4\phi)} \cos(r^2\sin 4\phi +2\phi -\xi +
	\lambda)\; . 
	\label{eq:spm-mean}
\end{equation}
Next, we calculate
\begin{equation}
	\langle\alpha|\hat{x}_{\lambda}'|\alpha\rangle = \langle\alpha
	e^{2i\theta}|\hat{x}_{\lambda}'|\alpha e^{2i\theta}\rangle  
	\label{eq:lambda-prime}
\end{equation}
for $\alpha = r\in\mathbb{R}$. This leads to $\cos(r^2\sin 4\phi +
2\phi - 2\theta + \lambda) = \cos(r^2\sin 4\phi + 2\phi + \lambda)$,
or 
\begin{equation}
	\lambda = \theta - r^2\sin 4\phi -2\phi \; .
	\label{eq:lambda-spm}
\end{equation}
Note that we have evaluated $\langle\hat{x}_{\lambda}'\rangle$ with
$\phi\rightarrow 2\phi$ since the {\sc spm} effect occurs {\em twice}
in the parity gate. We see that the value of $\lambda$ changes
dramatically, since it now depends on the magnitude $r$ of the
coherent state in the bus mode ($\lambda$ will generally be a large
multiple of $2\pi$). This means that $\lambda$ must be set with
extremely high precision. For the sake of the argument, we assume here
that $\lambda$ can be set exactly as in Eq.~(\ref{eq:lambda-spm}). 

The next step in the analysis of the weak nonlinear parity gate with
{\sc spm} is to evaluate the variance with respect to the three
different coherent bus states. To this end, we calculate
$\langle(\hat{x}_{\lambda}')^2\rangle$, again with $\phi\rightarrow
2\phi$. The expectation value of the operator $(\hat{x}_{\lambda}')^2$ is 
\begin{widetext}
\begin{equation}
	\langle(\hat{x}_{\lambda}')^2\rangle = \frac{1}{2} + r^2 + r^2 e^{-r^2(1-\cos 8\phi)} \cos \left( 2\lambda-2\xi + 8\phi +r^2\sin 8\phi \right) .
	\label{eq:spm-variance}
\end{equation}
We use the value of $\lambda$ given in Eq.~(\ref{eq:lambda-spm}) and $\theta = 2\phi$, and we make the approximations $\cos x = 1-x^2/2$ and $\sin x = x - x^3/6$. The variance of $\hat{x}_{\lambda}'$ for the three coherent states $|\alpha\rangle$, $|\alpha e^{i\theta}\rangle$, and $|\alpha e^{2i\theta}\rangle$ then becomes
\begin{eqnarray}
	(\Delta\hat{x}_{\lambda}')^2_0 &=& \frac{1}{2} + r^2 + r^2\, e^{-8r^2\theta^2} \cos(4\theta - 8r^2\theta^3) - 2r^2\, e^{-4r^2\theta^2} \cos^2\theta\; , \\
	(\Delta\hat{x}_{\lambda}')^2_{\theta} &=& \frac{1}{2} + r^2 + r^2\, e^{-8r^2\theta^2} \cos(2\theta - 8r^2\theta^3) - 2r^2\, e^{-4r^2\theta^2}\; , \\
	(\Delta\hat{x}_{\lambda}')^2_{2\theta} &=& \frac{1}{2} + r^2 + r^2\, e^{-8r^2\theta^2} \cos(8r^2\theta^3) - 2r^2\, e^{-4r^2\theta^2} \cos^2\theta\; ,
	\label{eq:spm-variances}
\end{eqnarray}
\end{widetext}
It is clear that the variances for the two different even parity
states are not identical, and the measurement will generally cause an
outcome-dependent rotation in the even parity subspace. However, since
the measurement outcome is known (it is needed to determine the
corrective phase shift), the state of the qubits remains pure. The
gate introduces a so-called {\em tilting} error, which can be absorbed
in an adaptive strategy for cluster state generation
\cite{campbell07}. The corrective phase shift needed for the gate
operation can be calculated using Eq.~(\ref{eq:generating}) (see also
Rohde et al. \cite{rohde07}).

\section{Resolution criterium}

\begin{figure}[b]
\begin{center}
\begin{psfrags}
 \psfrag{r}[t]{$r$}
 \psfrag{t}{$\theta$}
 \psfrag{S}[r]{$S$}
 \psfrag{0}{0}
 \psfrag{2}{2}
 \psfrag{4}{4}
 \psfrag{6}{6}
 \psfrag{8}{8}
 \psfrag{10}{10}
 \psfrag{0.02}[b]{0.02}
 \psfrag{0.04}[b]{0.04}
 \psfrag{0.06}[b]{0.06}
 \psfrag{0.08}[b]{0.08}
 \psfrag{0.1}{0.1}
 \psfrag{-100}{~$\,\;-1$}
 \psfrag{-200}{~$\,\;-2$}
 \psfrag{-300}{~$\,\;-3$}
  \includegraphics[width=85mm]{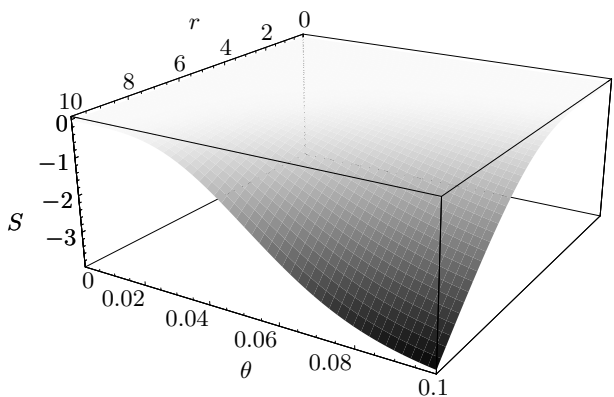}
 \end{psfrags}
 \end{center}
 \caption{The difference $S$ between the separation of the means
 $|\langle\hat{x}_{\lambda}'\rangle_e -
 \langle\hat{x}_{\lambda}'\rangle_o|$ and the sum of the variances
 $(\Delta \hat{x}_{\lambda}')_e + (\Delta \hat{x}_{\lambda}')_o$ in
 arbitrary units. This quantity is strictly negative, which means that
 a quadrature measurement can never distinguish between even and odd
 parity when the self-phase modulation is half the size of the
 cross-phase modulation.} 
 \label{fig:req}
\end{figure}

\noindent
Given the setup discussed in the previous section, we now ask what is
the parameter range that provides good distinguishability between even
and odd parity projections. Distinguishability of the parities
requires that the means of the two distributions (of even and odd
parity) must be larger than the sum of the variances of the parity
distributions: 
\begin{equation}
	|\langle\hat{x}_{\lambda}'\rangle_e - \langle\hat{x}_{\lambda}'\rangle_o| > (\Delta \hat{x}_{\lambda}')_o + \max(\Delta \hat{x}_{\lambda}')_e  \; .
	\label{eq:req}
\end{equation}
where the subscripts $e$ and $o$ denote evaluation with respect to the
even or odd parity distribution, respectively, and $\max(\Delta
\hat{x}_{\lambda}')_e$ indicates that we need to evaluate the
inequality using the largest variance from the pair
$(\Delta\hat{x}_{\lambda}')^2_0$ and
$(\Delta\hat{x}_{\lambda}')^2_{2\theta}$. This is the resolution
criterium. Using the means and variances of the previous section
(where $\theta = 2\phi$), we find that the requirement in
Eq.~(\ref{eq:req}) is never satisfied. Since    
\begin{equation}
	(\Delta \hat{x}_{\lambda}')_o + \max(\Delta
	\hat{x}_{\lambda}')_e \geq 2 \min(\Delta \hat{x}_{\lambda}')_j
	\; 
	\label{eq:min}
\end{equation}
with $j\in\{ 0,\theta,2\theta\}$, we can instead evaluate 
\begin{equation}
	|\langle\hat{x}_{\lambda}'\rangle_e -
         \langle\hat{x}_{\lambda}'\rangle_o|^2 > 4 \min(\Delta
         \hat{x}_{\lambda}')^2_j 
	\label{eq:minresolution}
\end{equation}
to show that the two probability distributions are never resolved in a
measurement of the $\hat{x}_{\lambda}$ quadrature. We plot the
difference $S$ between the mean separation and this variance in
Fig.~\ref{fig:req} (for $r$ and $\theta$ in arbitrary units), and see
that this function never exceeds zero. This means that for the weak
nonlinear parity gate the {\sc spm} must be negligable. 

The {\sc spm} is a non-Gaussian operation, and {\em a priori} there is
no reason for the two probability distributions to obey the resolution
criterium in Eq.~(\ref{eq:req}). The {\sc spm} may change the
distributions in such a way that the variance becomes large, but the
distributions still give two distinct peaks. Such a distribution is
said to be {\em leptokurtic}, with a positive kurtosis. On the other
hand, in a {\em platykurtic} distribution, where the kurtosis is
negative, the values tend to lie further away from the mean. In this
case the resolution criterium based on the variance is valid. We also
need to consider the skewness, or asymmetry, of the distributions. For
highly asymmetric distributions the variance will not be a good
resolution criterium. In order to convince ourselves that the
criterium is valid, we calculate the skewness and the kurtosis of the
probability distributions.  

\begin{figure}[b]
\begin{center}
\begin{psfrags}
 \psfrag{0}{0}
 \psfrag{10}{10}
 \psfrag{20}{20}
 \psfrag{30}{30}
 \psfrag{40}{40}
 \psfrag{50}{50}
 \psfrag{0.02}{0.02}
 \psfrag{0.04}{0.04}
 \psfrag{0.06}{0.06}
 \psfrag{0.08}{0.08}
 \psfrag{0.1}{0.1}
 \psfrag{a}{$r$}
 \psfrag{b}{$\theta$} 
  \includegraphics[width=85mm]{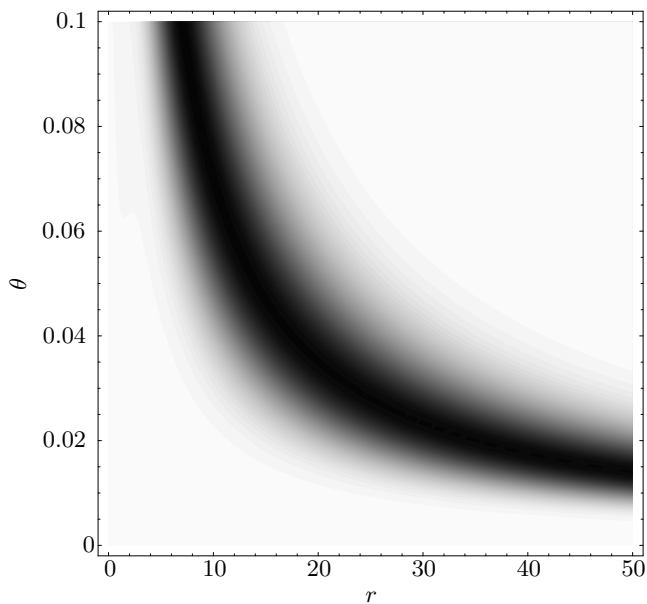}
 \end{psfrags}
 \end{center}
 \caption{The skewness is close to zero everywhere (darker equals more negative), except around a curve $r^2 \approx (2\theta^2)^{-1}$, where $\gamma_1 < 0$. In the non-phase modulated parity gate regime of operation $r \gtrsim \theta^{-2}$ the skewness is negligable.}
 \label{fig:skewness}
\end{figure}

\begin{figure}[b]
\begin{center}
\begin{psfrags}
 \psfrag{0}{0}
 \psfrag{10}{10}
 \psfrag{20}{20}
 \psfrag{30}{30}
 \psfrag{40}{40}
 \psfrag{50}{50}
 \psfrag{0.02}{0.02}
 \psfrag{0.04}{0.04}
 \psfrag{0.06}{0.06}
 \psfrag{0.08}{0.08}
 \psfrag{0.1}{0.1}
 \psfrag{a}{$r$}
 \psfrag{b}{$\theta$} 
  \includegraphics[width=85mm]{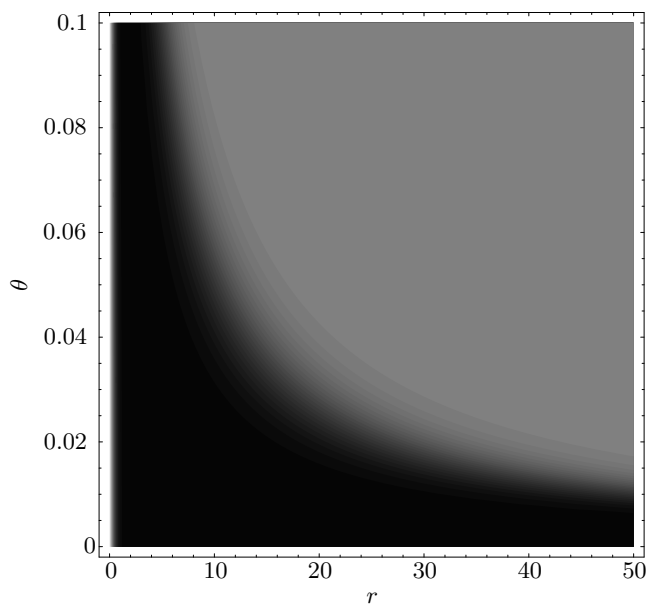}
 \end{psfrags}
 \end{center}
 \caption{The kurtosis in the parameter regime $r\gtrsim
 (2\theta)^{-2}$ is approximately $-1.5$ (darker is more negative),
 which means that the distribution is platykurtic. The values of the
 probability distribution tend to lie further away from the mean.} 
 \label{fig:kurtosis}
\end{figure}

The skewness $\gamma_1$ of a probability distribution can be expressed in terms of the third moment about the mean $\mu_3$ and the standard deviation $\sigma = \Delta \hat{x}_{\lambda}'$: 
\begin{equation}
	\gamma_1 = \frac{\mu_3}{(\Delta \hat{x}_{\lambda}')^3} \quad\text{and}\quad 
	\mu_3 = \langle (\hat{x}_{\lambda}' - \langle\hat{x}_{\lambda}'\rangle)^3 \rangle\; .
	\label{eq:skewness}
\end{equation}
A positive skewness implied an elongated tail towards the positive end of the parameter space. The third moment about the mean can be written in terms of expectation values of the higher order moments of $\hat{x}_{\lambda}'$:
\begin{equation}
	\mu_3 = \langle(\hat{x}_{\lambda}')^3\rangle -3 \langle(\hat{x}_{\lambda}')^2\rangle \langle\hat{x}_{\lambda}'\rangle + 2 \langle\hat{x}_{\lambda}'\rangle^3\; .
	\label{eq:mu3}
\end{equation}
Similarly, the kurtosis $\gamma_2$ of a probability distribution can be defined in terms of the fourth moment about the mean $\mu_4$ and the standard deviation $\sigma = \Delta \hat{x}_{\lambda}'$: 
\begin{equation}
	\gamma_2 = \frac{\mu_4}{(\Delta \hat{x}_{\lambda}')^4} - 3 \quad\text{and}\quad 
	\mu_4 = \langle (\hat{x}_{\lambda}' - \langle\hat{x}_{\lambda}'\rangle)^4 \rangle\; .
	\label{eq:kurtosis}
\end{equation}
In terms of the higher order moments of $\hat{x}_{\lambda}'$, we can write $\mu_4$ as
\begin{equation}\nonumber
	\mu_4 = \langle(\hat{x}_{\lambda}')^4\rangle - 4 \langle(\hat{x}_{\lambda}')^3\rangle \langle\hat{x}_{\lambda}'\rangle + 6 \langle(\hat{x}_{\lambda}')^2\rangle \langle\hat{x}_{\lambda}'\rangle^2 - 3 \langle\hat{x}_{\lambda}'\rangle^3\; .
	\label{eq:mu4}
\end{equation}
We evaluate the expectation values $\langle(\hat{x}_{\lambda}')^3\rangle$ and  $\langle(\hat{x}_{\lambda}')^4\rangle$ with respect to coherent states $|\alpha\rangle$, where $\alpha = r\, e^{i\xi}$:
\begin{widetext}
\begin{eqnarray}
	\langle(\hat{x}_{\lambda}')^3\rangle &=& \frac{r^3}{\sqrt{2}} e^{-r^2(1-\cos 12\phi)} \cos(3\lambda-3\xi+18\phi+r^2\sin 12\phi) + \frac{3(r^3+r)}{\sqrt{2}} e^{-r^2(1-\cos 4\phi)} \cos(\lambda-\xi+6\phi+r^2\sin 4\phi) \cr
	\langle(\hat{x}_{\lambda}')^4\rangle &=& \frac{r^4}{2} e^{-r^2(1-\cos 16\phi)} \cos(4\lambda-4\xi+32\phi+r^2\sin 16\phi) \cr && + (2r^4+3r^2) e^{-r^2(1-\cos 8\phi)} \cos(2\lambda-2\xi+16\phi+r^2\sin 8\phi) + \frac{3}{2} r^4 + 3r^2 + \frac{3}{4}  .
	\label{eq:moments3and4}
\end{eqnarray}
\end{widetext}
These expressions are exact. Using the value $\lambda = \theta - 2\phi
- r^2\sin 4\phi$ and $\theta=2\phi$, we plot the skewness and kurtosis
as a function of $r$ and $\theta$ in Figs.~\ref{fig:skewness} and
\ref{fig:kurtosis}. In the parameter regime $r\theta^2$ where the
non-phase modulated parity gate operates, the skewness is practically
zero and the kurtosis is negative. This indicates that the effect of
{\sc spm} indeed destroys the distinguishability between the
probability distributions for even and odd parity. 

\section{Squeezing}

Finally, we may salvage the weak nonlinear parity gate by employing
quadrature squeezing of the bus mode after the effects of {\sc
  spm}. The squeezing must be in the $\hat{x}_{\lambda}$ direction
such that the peaks in the marginal probability distribution $P$
become narower. Typically, when $\zeta$ is the squeezing parameter,
the variance is reduced by a factor $\exp(-\zeta)$ and the resolution
criterium becomes 
\begin{equation}
 |\langle\hat{x}'_{\lambda}\rangle_e -
  \langle\hat{x}'_{\lambda}\rangle_o|\, e^{\zeta} > (\Delta
  \hat{x}'_{\lambda})_o + \max(\Delta \hat{x}'_{\lambda})_e\, .
\end{equation}
The down-side of using squeezing is that if $\zeta$ is fairly large
and $\lambda$ is not chosen sufficiently accurate (which, we have
seen, is rather difficult), it may exacerbate the indistinguishability
of the peaks. Squeezing in the conjugate quadrature will result in an
enhancement of the variance with a factor $\exp(\zeta)$. In a rotated
frame $\hat{x}_{\varphi} = \cos\varphi\,\hat{x} +
\sin\varphi\,\hat{x}_{\pi/2}$ the variance of a squeezed coherent
state becomes
\begin{equation}
 (\Delta \hat{x}_{\varphi})^2 = \cos^2\varphi\, e^{-2\zeta} +
 \sin^2\varphi\, e^{2\zeta}\, .
\end{equation}
If the use of squeezing is to outperform the setup without squeezing,
the offset in the rotation angle must be smaller than $\tan\varphi =
\exp(-\zeta)$.

\section{Conclusions}

Typically, weak nonlinear gates that do not propose to use EIT
materials involve a self-phase modulation of the bus mode. In typical
nonlinear materials this effect is half the size of the cross-phase
modulation. Also, the cross-phase modulation (and hence the self-phase
modulation) may not be easily switched in the course of operating the
gate. This means that the gate must be redesigned to operate with two
successive identical phase shifts. In this case the self-phase
modulation no longer cancels, as it cannot be switched in sign
independently from the cross-phase modulation.

In this paper, it was shown that a typical self-phase modulation half
the size of the cross-phase modulation will remove the
distinguishability of the two parity measurement outcomes. This
implementation of the gate will therefore not work without extra
quadrature squeezing, which in turn presents difficulties regarding
the tuning of the quadrature.   The Wigner function and the marginal
quadrature probability distribution for the bus state was constructed
in terms of easily calculable generating functions.

\section*{Acknowledgments}

The author wishes to thank the members of the Quantum and
Nano-Technology Group (www.qunat.org) for stimulating discussions.
Part of this research was carried out for the QIP IRC www.qipirc.org
(GR/S82176/01).


\begin{thebibliography}{17}
\expandafter\ifx\csname natexlab\endcsname\relax\def\natexlab#1{#1}\fi
\expandafter\ifx\csname bibnamefont\endcsname\relax
  \def\bibnamefont#1{#1}\fi
\expandafter\ifx\csname bibfnamefont\endcsname\relax
  \def\bibfnamefont#1{#1}\fi
\expandafter\ifx\csname citenamefont\endcsname\relax
  \def\citenamefont#1{#1}\fi
\expandafter\ifx\csname url\endcsname\relax
  \def\url#1{\texttt{#1}}\fi
\expandafter\ifx\csname urlprefix\endcsname\relax\def\urlprefix{URL }\fi
\providecommand{\bibinfo}[2]{#2}
\providecommand{\eprint}[2][]{\url{#2}}

\bibitem[{\citenamefont{Knill et~al.}(2001)\citenamefont{Knill, Laflamme, and
  Milburn}}]{knill01}
\bibinfo{author}{\bibfnamefont{E.}~\bibnamefont{Knill}},
  \bibinfo{author}{\bibfnamefont{R.}~\bibnamefont{Laflamme}}, \bibnamefont{and}
  \bibinfo{author}{\bibfnamefont{G.~J.} \bibnamefont{Milburn}},
  \bibinfo{journal}{Nature} \textbf{\bibinfo{volume}{409}}, \bibinfo{pages}{46}
  (\bibinfo{year}{2001}).

\bibitem[{\citenamefont{Kok et~al.}(2007)\citenamefont{Kok, Munro, Nemoto,
  Ralph, Dowling, and Milburn}}]{kok07}
\bibinfo{author}{\bibfnamefont{P.}~\bibnamefont{Kok}},
  \bibinfo{author}{\bibfnamefont{W.~J.} \bibnamefont{Munro}},
  \bibinfo{author}{\bibfnamefont{K.}~\bibnamefont{Nemoto}},
  \bibinfo{author}{\bibfnamefont{T.~C.} \bibnamefont{Ralph}},
  \bibinfo{author}{\bibfnamefont{J.~P.} \bibnamefont{Dowling}},
  \bibnamefont{and} \bibinfo{author}{\bibfnamefont{G.~J.}
  \bibnamefont{Milburn}}, \bibinfo{journal}{Rev. Mod. Phys.}
  \textbf{\bibinfo{volume}{79}}, \bibinfo{pages}{135} (\bibinfo{year}{2007}).

\bibitem[{\citenamefont{Barrett and Kok}(2005)}]{barrett05}
\bibinfo{author}{\bibfnamefont{S.~D.} \bibnamefont{Barrett}} \bibnamefont{and}
  \bibinfo{author}{\bibfnamefont{P.}~\bibnamefont{Kok}},
  \bibinfo{journal}{Phys. Rev. A} \textbf{\bibinfo{volume}{72}},
  \bibinfo{pages}{060310} (\bibinfo{year}{2005}).

\bibitem[{\citenamefont{Lim et~al.}(2005)\citenamefont{Lim, Beige, and
  Kwek}}]{lim05}
\bibinfo{author}{\bibfnamefont{Y.~L.} \bibnamefont{Lim}},
  \bibinfo{author}{\bibfnamefont{A.}~\bibnamefont{Beige}}, \bibnamefont{and}
  \bibinfo{author}{\bibfnamefont{L.~C.} \bibnamefont{Kwek}},
  \bibinfo{journal}{Physical Review Letters} \textbf{\bibinfo{volume}{95}},
  \bibinfo{pages}{030505} (\bibinfo{year}{2005}).

\bibitem[{\citenamefont{Lim et~al.}(2006)\citenamefont{Lim, Barrett, Beige,
  Kok, and Kwek}}]{lim06}
\bibinfo{author}{\bibfnamefont{Y.~L.} \bibnamefont{Lim}},
  \bibinfo{author}{\bibfnamefont{S.~D.} \bibnamefont{Barrett}},
  \bibinfo{author}{\bibfnamefont{A.}~\bibnamefont{Beige}},
  \bibinfo{author}{\bibfnamefont{P.}~\bibnamefont{Kok}}, \bibnamefont{and}
  \bibinfo{author}{\bibfnamefont{L.~C.} \bibnamefont{Kwek}},
  \bibinfo{journal}{Physical Review A} \textbf{\bibinfo{volume}{73}},
  \bibinfo{pages}{012304} (\bibinfo{year}{2006}).

\bibitem[{\citenamefont{Franson et~al.}(2004)\citenamefont{Franson, Jacobs, and
  Pittman}}]{franson04}
\bibinfo{author}{\bibfnamefont{J.~D.} \bibnamefont{Franson}},
  \bibinfo{author}{\bibfnamefont{B.~C.} \bibnamefont{Jacobs}},
  \bibnamefont{and} \bibinfo{author}{\bibfnamefont{T.~B.}
  \bibnamefont{Pittman}}, \bibinfo{journal}{Phys. Rev. A}
  \textbf{\bibinfo{volume}{70}}, \bibinfo{pages}{062302}
  (\bibinfo{year}{2004}).

\bibitem[{\citenamefont{Barrett et~al.}(2005)\citenamefont{Barrett, Kok,
  Nemoto, Beausoleil, Munro, and Spiller}}]{barrett05b}
\bibinfo{author}{\bibfnamefont{S.~D.} \bibnamefont{Barrett}},
  \bibinfo{author}{\bibfnamefont{P.}~\bibnamefont{Kok}},
  \bibinfo{author}{\bibfnamefont{K.}~\bibnamefont{Nemoto}},
  \bibinfo{author}{\bibfnamefont{R.~G.} \bibnamefont{Beausoleil}},
  \bibinfo{author}{\bibfnamefont{W.~J.} \bibnamefont{Munro}}, \bibnamefont{and}
  \bibinfo{author}{\bibfnamefont{T.~P.} \bibnamefont{Spiller}},
  \bibinfo{journal}{Phys. Rev. A} \textbf{\bibinfo{volume}{72}},
  \bibinfo{pages}{060302} (\bibinfo{year}{2005}).

\bibitem[{\citenamefont{Nemoto and Munro}(2004)}]{nemoto04}
\bibinfo{author}{\bibfnamefont{K.}~\bibnamefont{Nemoto}} \bibnamefont{and}
  \bibinfo{author}{\bibfnamefont{W.~J.} \bibnamefont{Munro}},
  \bibinfo{journal}{Phys. Rev. Lett.} \textbf{\bibinfo{volume}{93}},
  \bibinfo{pages}{250502} (\bibinfo{year}{2004}).

\bibitem[{\citenamefont{Spiller et~al.}(2006)\citenamefont{Spiller, Nemoto,
  Braunstein, Munro, van Loock, and Milburn}}]{spiller06}
\bibinfo{author}{\bibfnamefont{T.~P.} \bibnamefont{Spiller}},
  \bibinfo{author}{\bibfnamefont{K.}~\bibnamefont{Nemoto}},
  \bibinfo{author}{\bibfnamefont{S.~L.} \bibnamefont{Braunstein}},
  \bibinfo{author}{\bibfnamefont{W.~J.} \bibnamefont{Munro}},
  \bibinfo{author}{\bibfnamefont{P.}~\bibnamefont{van Loock}},
  \bibnamefont{and} \bibinfo{author}{\bibfnamefont{G.~J.}
  \bibnamefont{Milburn}}, \bibinfo{journal}{New J. Phys.}
  \textbf{\bibinfo{volume}{8}}, \bibinfo{pages}{30} (\bibinfo{year}{2006}).

\bibitem[{\citenamefont{Schmidt and Imamoglu}(1996)}]{schmidt96}
\bibinfo{author}{\bibfnamefont{H.}~\bibnamefont{Schmidt}} \bibnamefont{and}
  \bibinfo{author}{\bibfnamefont{A.}~\bibnamefont{Imamoglu}},
  \bibinfo{journal}{Opt. Lett.} \textbf{\bibinfo{volume}{21}},
  \bibinfo{pages}{1936} (\bibinfo{year}{1996}).

\bibitem[{\citenamefont{Chiao et~al.}(1966)\citenamefont{Chiao, Kelley, and
  Garmire}}]{chiao66}
\bibinfo{author}{\bibfnamefont{R.~Y.} \bibnamefont{Chiao}},
  \bibinfo{author}{\bibfnamefont{P.~L.} \bibnamefont{Kelley}},
  \bibnamefont{and} \bibinfo{author}{\bibfnamefont{E.}~\bibnamefont{Garmire}},
  \bibinfo{journal}{Phys. Rev. Lett.} \textbf{\bibinfo{volume}{17}},
  \bibinfo{pages}{1158} (\bibinfo{year}{1966}).

\bibitem[{\citenamefont{Boyd}(2003)}]{boyd03}
\bibinfo{author}{\bibfnamefont{R.~W.} \bibnamefont{Boyd}},
  \emph{\bibinfo{title}{Nonlinear optics}} (\bibinfo{publisher}{Academic
  Press}, \bibinfo{year}{2003}), \bibinfo{edition}{2nd} ed.

\bibitem[{\citenamefont{Browne and Rudolph}(2005)}]{browne05}
\bibinfo{author}{\bibfnamefont{D.~E.} \bibnamefont{Browne}} \bibnamefont{and}
  \bibinfo{author}{\bibfnamefont{T.}~\bibnamefont{Rudolph}},
  \bibinfo{journal}{Phys. Rev. Lett.} \textbf{\bibinfo{volume}{95}},
  \bibinfo{pages}{010501} (\bibinfo{year}{2005}).

\bibitem[{\citenamefont{Barrett and Milburn}(2006)}]{barrett06}
\bibinfo{author}{\bibfnamefont{S.~D.} \bibnamefont{Barrett}} \bibnamefont{and}
  \bibinfo{author}{\bibfnamefont{G.~J.} \bibnamefont{Milburn}},
  \bibinfo{journal}{Phys. Rev. A} \textbf{\bibinfo{volume}{74}},
  \bibinfo{pages}{060302} (\bibinfo{year}{2006}).

\bibitem[{\citenamefont{Barnett and Radmore}(1997)}]{barnett}
\bibinfo{author}{\bibfnamefont{S.~M.} \bibnamefont{Barnett}} \bibnamefont{and}
  \bibinfo{author}{\bibfnamefont{P.~M.} \bibnamefont{Radmore}},
  \emph{\bibinfo{title}{Methods in Theoretical Quantum Optics}}
  (\bibinfo{publisher}{Oxford University Press}, \bibinfo{year}{1997}).

\bibitem[{\citenamefont{Campbell et~al.}(2007)\citenamefont{Campbell,
  Fitzsimons, Benjamin, and Kok}}]{campbell07}
\bibinfo{author}{\bibfnamefont{E.~T.} \bibnamefont{Campbell}},
  \bibinfo{author}{\bibfnamefont{J.}~\bibnamefont{Fitzsimons}},
  \bibinfo{author}{\bibfnamefont{S.~C.} \bibnamefont{Benjamin}},
  \bibnamefont{and} \bibinfo{author}{\bibfnamefont{P.}~\bibnamefont{Kok}},
  \bibinfo{journal}{Phys. Rev. A} \textbf{\bibinfo{volume}{75}},
  \bibinfo{pages}{042303} (\bibinfo{year}{2007}).

\bibitem[{\citenamefont{Rohde et~al.}(2007)\citenamefont{Rohde, Munro, Ralph,
  van Loock, and Nemoto}}]{rohde07}
\bibinfo{author}{\bibfnamefont{P.~P.} \bibnamefont{Rohde}},
  \bibinfo{author}{\bibfnamefont{W.~J.} \bibnamefont{Munro}},
  \bibinfo{author}{\bibfnamefont{T.~C.} \bibnamefont{Ralph}},
  \bibinfo{author}{\bibfnamefont{P.}~\bibnamefont{van Loock}},
  \bibnamefont{and} \bibinfo{author}{\bibfnamefont{K.}~\bibnamefont{Nemoto}},
  \bibinfo{journal}{arXiv:0705.4522 [quant-ph]}  (\bibinfo{year}{2007}).

\end{thebibliography}
\end{document}